\documentclass[aip,amsmath,amssymb,reprint]{revtex4-1}

\usepackage{comment}
\usepackage[none]{hyphenat}
\usepackage[normalem]{ulem}
\usepackage{xcolor}

\usepackage{amsmath}
\usepackage{amsfonts}
\usepackage{amssymb}
\usepackage{graphicx}
\usepackage{caption}
\usepackage{subcaption}
\usepackage{mathtools}
\usepackage{natbib}
\usepackage[capitalise]{cleveref}

\makeatletter
\def\@email#1#2{
 \endgroup
 \patchcmd{\titleblock@produce}
  {\frontmatter@RRAPformat}
  {\frontmatter@RRAPformat{\produce@RRAP{*#1\href{mailto:#2}{#2}}}\frontmatter@RRAPformat}
  {}{}
}
\makeatother

\begin{document}

\title[Symmetry breaking of two-layer dipoles]{Symmetry breaking and nonlinear transformation of two-layer eastward propagating dipoles}
\author{Matthew. N. Crowe}
\affiliation{
School of Mathematics, Statistics and Physics, Newcastle University, Newcastle upon Tyne, NE1 7RU, UK.
}
\affiliation{
Department of Mathematics, University College London, London, WC1E 6BT, UK.
}

\author{Georgi. G. Sutyrin}
\affiliation{Graduate School of Oceanography, University of Rhode Island, Narragansett 02882, USA.}

\date{\today}

\begin{abstract}
We study the evolution of eastward propagating dipoles (modons) in a two-layer quasi-geostrophic $\beta$-plane model using high-resolution numerical simulations. 
Various combinations of background gradients of potential vorticity in the upper and lower layer (which may include sloping topography) shed light on the recently discovered breakdown mechanisms and rich dynamics of dipolar vortices.
Owing to the $\beta$-effect in the upper layer with active dipolar vortices, the symmetry of dipole flow breaks due to an exponentially-growing, rotating, asymmetric mode of linear instability associated with Rossby wave radiation. 
Further nonlinear transformation is found to consist of two phases: fast partner separation, resulting in a deceleration of the eastward drift, and subsequent slow separation with a saturated asymmetric mode accompanied by much weaker, shorter Rossby waves.
This weakly radiating phase---with pulsating partners and homogenized potential vorticity between the core and the separatrix---can be considered as a new type of long-lived dipole.
Conversely, when no $\beta$-effect is present in the upper layer, the dipoles remain nearly symmetric, even when the topographic $\beta$-effect is present in the active lower layer. In this case, the development of a weak asymmetric mode is related to a small meridional shift of the dipole center on the numerical grid.
\end{abstract}

\maketitle


Self-propagating dipoles, consisting of a closely packed pair of counter-rotating vortices, are commonly observed in nature \citep{FederovGinsburg,HughesMiller,NiEtAl} and have been extensively investigated across various theoretical studies \citep{NielsenRasmussen,LewekeEtAl, Crowe_Johnson_2024}.
In geophysical fluid dynamics and plasma physics, vortex dynamics are strongly affected by Rossby waves which arise from large-scale gradients in the ambient potential vorticity.
{These background gradients are caused by} the $\beta$-effect, topography and large-scale currents \citep{SokolovskiiVerron, BerloffSutyrin}.
Over several decades, many theoretical solutions describing eastward propagating dipoles (EPD) have been thought to remain steady, avoiding resonance with the Rossby waves propagating westward \citep{FlierlEtAl}. 
A new type of linear instability associated with EPD symmetry breaking and resulting in eventual dipole disintegration was recently identified numerically for an upper-ocean reduced-gravity model \citep[referred to here as DSB23]{DaviesEtAl_1}.
This work applies recent developments in numerical modelling to the classical problem of breakdown in dipolar vortices and highlights interesting new behaviours which have not previously been observed.

Families of steady propagating dipoles on the $f$-plane (i.e. with no $\beta$-effect) originate from the barotropic Lamb-Chaplygin solution to the 2D Euler equations \citep{Lamb_1932,Chaplygin,MeleshkoH94}.
While these solutions have long been known, the stability properties of the inviscid Lamb-Chaplygin dipole (LCD) are still not well understood \citep{Protas_2024}. Further, it has been shown that 2D perturbations to the LCD can grow exponentially in the presence of viscosity \citep{BrionEtAl}. 
Similarly, an analogue to the LCD on the $\beta$-plane, a 2D EPD referred to as a ``modon'' by \citet{Stern}, was found to be unstable by \citet{KiznerBerson}.
 
In the equivalent-barotropic (reduced-gravity) model of the upper ocean, a stationary dipole solution on the $\beta$-plane was obtained using the quasigeostrophic (QG) approximation by \citet{LarichevR76d}. 
Multiple attempts have failed to prove analytically that steady Larichev-Reznik EPD are stable \citep{Nycander}. 
In the limit $\beta \to 0$, these solutions can be considered as an analogue of 2D LCD on the $f$-plane with finite radius of deformation.
On the $\beta$-plane, such $f$-plane upper-layer dipoles {evolve} depending on their intensity and initial direction of propagation \citep{SutyrinEtAl}.
Another study with a focus on Larichev-Reznik {$\beta$-plane} solutions tilted away from the zonal direction found that their ability to adjust to steady EPD depends on the initial tilt and the value of $\beta$ 
\citep{NycanderI90,HesthavenEtAl}.
A general framework for Rossby wave generation and associated energy loss has been formulated recently, and applied to a range of dipolar vortex problems \citep{JohnsonC21, Crowe_Johnson_21, Crowe_Johnson_23}.

Both tilted and zonal Larichev-Reznik EPD were recently found to experience spontaneous symmetry breaking and instability \citep{DaviesEtAl_2}. 
The fastest growing normal modes were extracted by
both solving the initial-value problem and by direct calculation of the eigenvalue spectrum \citep{DaviesEtAl_3}.
The growth rate of the EPD instability was found to decrease with the value of $\beta$ for $\beta>\beta_c$ (see definition below).
These results were obtained by utilising modern numerical techniques and high spatial resolution, surpassing those used in previous studies.

However, the EPD evolution for smaller $\beta$ and the role of the lower layer feedback in the evolution and breakdown of modons remain unclear. Here, we will consider the long-time evolution of two-layer dipoles\citep{ReznikSutyrin,KiznerEtAl,Crowe_Johnson_2024} in a QG model with sloping bottom. 


We consider a two-layer rotating fluids on the $\beta$-plane, where the lower layer depth can change in the meridional ($y$) direction. 
Under the QG approximation, the governing equations are 
\begin{equation}\label{QG}
(\partial_t-c\partial_x)q_j + J[\psi_j,q_j] + \beta_j \partial_x\psi_j = 0,
\end{equation}
for streamfunction, $\psi_j$, and potential vorticity anomaly (PVA), $q_j$, in each layer where
\begin{equation}
q_j = \nabla^2 \psi_j - \frac{f_0^2}{g'h_j}(\psi_j-\psi_{3-j}), \hspace{0.5cm} j = 1, 2.
\end{equation}
Here $g'$ denotes the buoyancy difference between layers of thickness $h_j$.
In the upper layer $\beta_1=\beta$ is the meridional gradient of the Coriolis parameter, $f=f_0+\beta y$, while in the lower layer $\beta_2=\beta+(f_0/h_2)\alpha$ includes the constant topographic slope $\alpha$. Small-scale dissipation is not included in this model but is required for numerical simulations as described below. 
The equations are written in a Cartesian system with coordinates translating at a prescribed zonal ($x$) velocity, $c$, and $J[\phi,\varphi] = \phi_x\varphi_y - \phi_y\varphi_x$ denotes the Jacobian derivative.

The dimensionless variables 
\begin{equation}
T=tc/a,\, (X, Y)=(x, y)/a,\, Q_j=aq_j/c,\, \Psi_j=\psi_j/ac,
\end{equation}  
are used throughout this study.
We take the initial dipole radius to equal the Rossby radius for the upper layer, $a = R_1= \sqrt{g'h_1}/f_0$.
Throughout this work we consider both 1- and 2-layer cases.
For 1-layer cases, we take $\Psi_2 \equiv 0$, so the system reduces to the equivalent-barotropic problem, and denote them by `1L$\beta_1$'.
For the 2-layer cases, we set the lower depth to equal that of the upper layer, $h_2=h_1$, and denote them by `2L$\beta_1$:$\beta_2$'.


Steady propagating solutions are obtained by assuming $\partial_t = 0$ in \cref{QG} to give
\begin{equation}\label{eq5}
J[\Psi_j+Y, Q_j+(\beta_j/\beta_c) Y] = 0,
\end{equation}
where $\beta_c = c/a^2$. Therefore, the potential vorticity in each layer can be written as a function of the streamfunction as
\begin{equation}\label{eq6}
Q_j+(\beta_j/\beta_c) Y = \mathcal{F}_j(\Psi_j+Y),
\end{equation}
where the $\mathcal{F}_j$ are arbitrary (piece-wise differentiable) functions.
The linear form of $\mathcal{F}_j(z)=(\beta_j/\beta_c)\, z$ in the exterior domain is set by the requirement that vorticity and streamfunction perturbations decay towards
infinity and gives that $Q_j = (\beta_j/\beta_c)\,\Psi_j$ outside the vortex core, $X^2 + Y^2 > 1$.

Taking $\mathcal{F}_j$ to also be linear inside the vortex, $X^2 + Y^2 \leq 1$, gives a variety of solitary baroclinic eddy solutions. 
Here, we consider cases with an active vortex region in the upper layer only, assuming 
\begin{equation}\label{eq8}
\mathcal{F}_1(z)=-\kappa_1^2\, z \quad \mathrm{for} \quad X^2+Y^2 < 1, 
\end{equation}
where $\kappa_1$ represents an `internal wavenumber' for the vortex. The lower layer is referred to as `passive' as we assume that $\mathcal{F}_2(z)=(\beta_2/\beta_c)\, z$ everywhere. In polar coordinates, $(X,Y)=(r\cos\vartheta,r\sin\vartheta)$, a dipolar solution may be written as
\begin{equation}\label{modon}
\Psi_j = \Phi_j(r)\sin\vartheta,
\end{equation}
where the $\Phi_j(r)$ describe the radial vortex structure in each layer. An efficient semi-analytical method for finding fully nonlinear modon solutions in a multi-layer, quasi-geostrophic model with arbitrarily many layers was recently suggested by \citet{Crowe_Johnson_2024} using the Hankel tranform
and an expansion in terms of Zernike polynomials. The resulting system may be solved for $\Phi_j(r)$ with the values of the $\kappa_j$ appearing as eigenvalues in a linear algebra problem. Since we consider modons with a passive lower layer, only the value of $\kappa_1$ needs to be determined. Note that these solutions are the same as a `Regular Modon with one Interior Domain' as discussed by \citet{KiznerEtAl}.

\begin{table*}
\centering
\begin{tabular}{ ccccc } 
 Case & Two-layer flat bottom & Two-layer sloping botom & Upper layer \\
 Notation & 2L1:1 & 2L1:0 & 1L1  \\
 $\beta_2/\beta_c$ & 1 & 0 & 0 \\
 $U_{1max}$ & 5 & 5 & 5.3 \\
 $Q_{1max}$ & 20 & 19 & 21 \\
 $T_c$ & 275 & 330 & 350 \\
 $\sigma$ & 0.1 & 0.07 & 0.06 \\
\end{tabular}
\caption{Summary of the $\beta$-plane cases.}
\label{tab:tab1}
\end{table*}
\begin{figure*}
	\centering
	\begin{subfigure}[b]{0.24\textwidth}
        \caption{\hfill\,\vspace{-2pt}}
	\centering
	\includegraphics[trim={0.4cm 0.2cm 0.8cm 0.7cm},clip,width=\textwidth]{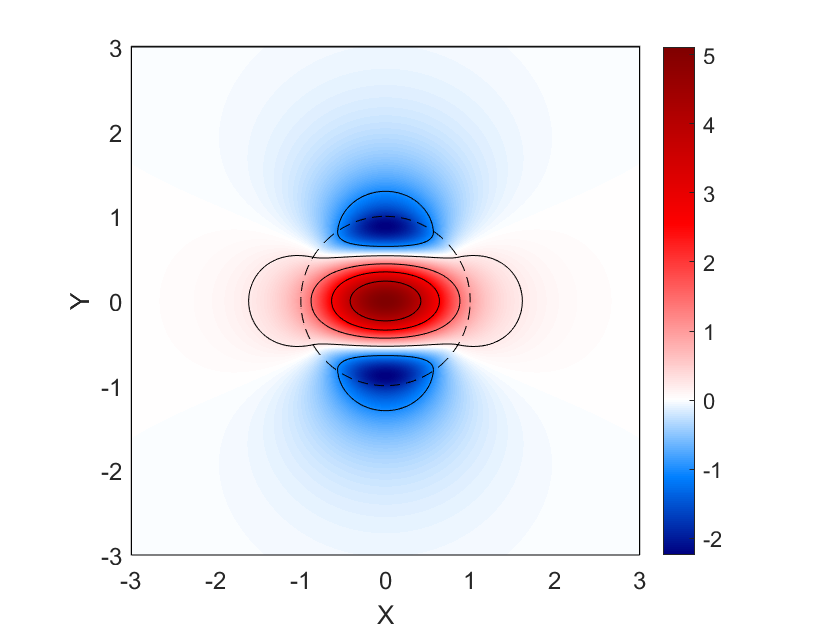}
	\end{subfigure}
	\begin{subfigure}[b]{0.24\textwidth}
        \caption{\hfill\,\vspace{-2pt}}
	\centering
	\includegraphics[trim={0.4cm 0.2cm 0.8cm 0.7cm},clip,width=\textwidth]{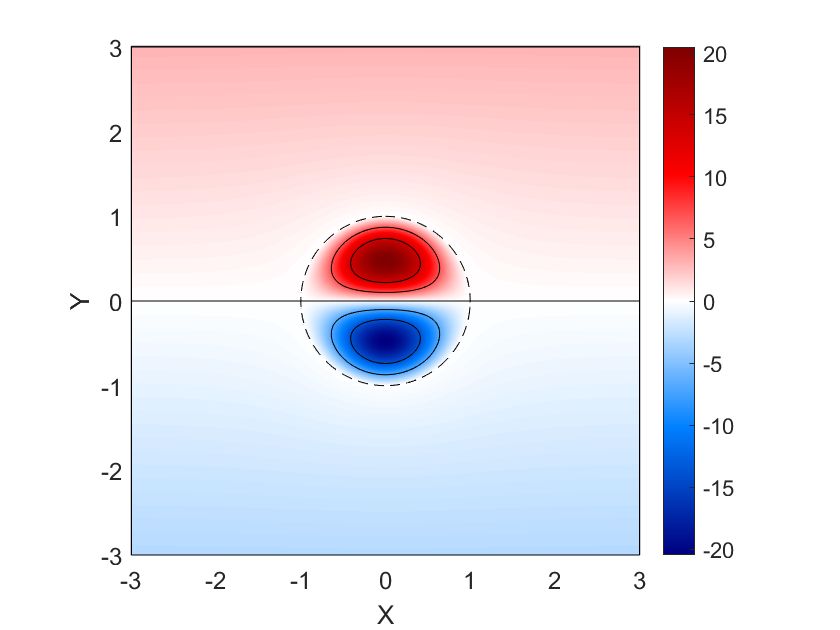}
	\end{subfigure}
	\begin{subfigure}[b]{0.24\textwidth}
	\caption{\hfill\,\vspace{-2pt}}
	\centering
	\includegraphics[trim={0.4cm 0.2cm 0.8cm 0.7cm},clip,width=\textwidth]{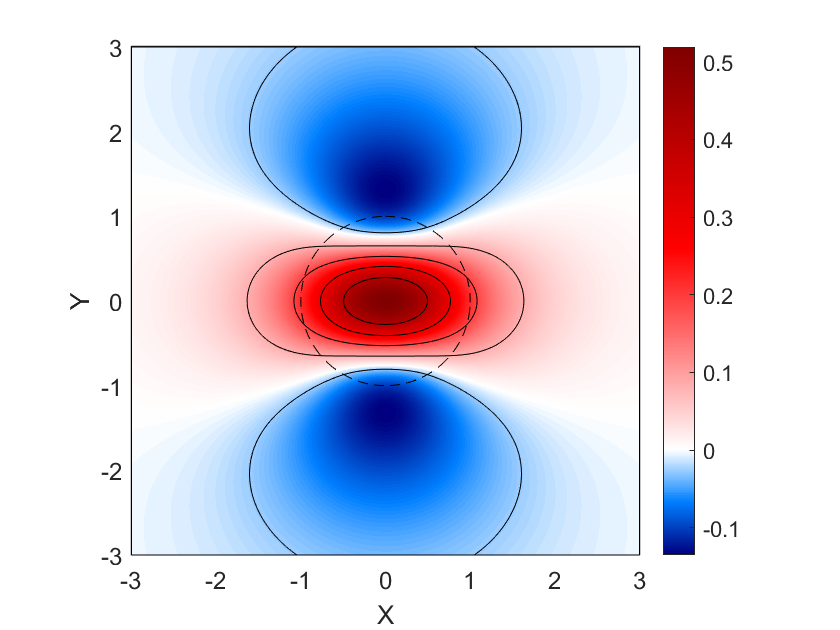}
	\end{subfigure}
	\begin{subfigure}[b]{0.24\textwidth}
 	\caption{\hfill\,\vspace{-2pt}}
	\centering
	\includegraphics[trim={0.4cm 0.2cm 0.8cm 0.7cm},clip,width=\textwidth]{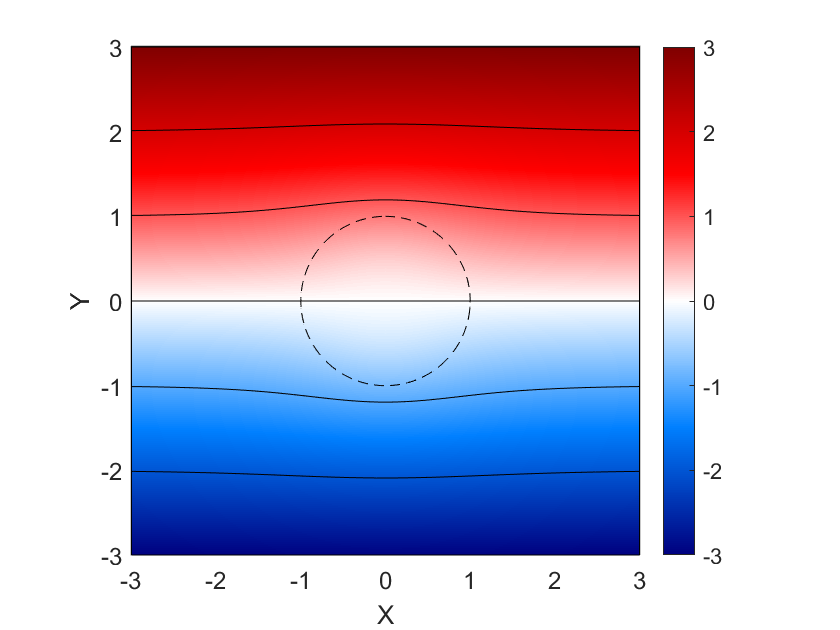}
	\end{subfigure}
	\caption{Initial fields for the case 2L1:1. We plot: upper zonal velocity, $-\partial\Psi_1/\partial Y$ (a); upper PV $Q_1+Y$ (b); deep zonal velocity $-\partial\Psi_2/\partial Y$ (c); and deep PV $Q_2+Y$ (d).}
    \label{fig:Fig_1}
 \end{figure*}
\begin{figure*}
	\centering
	\begin{subfigure}[b]{0.24\textwidth}
	\caption{\hfill\,\vspace{-2pt}}
        \centering
	\includegraphics[trim={0cm 0.1cm 0.8cm 0.2cm},clip,width=\textwidth]{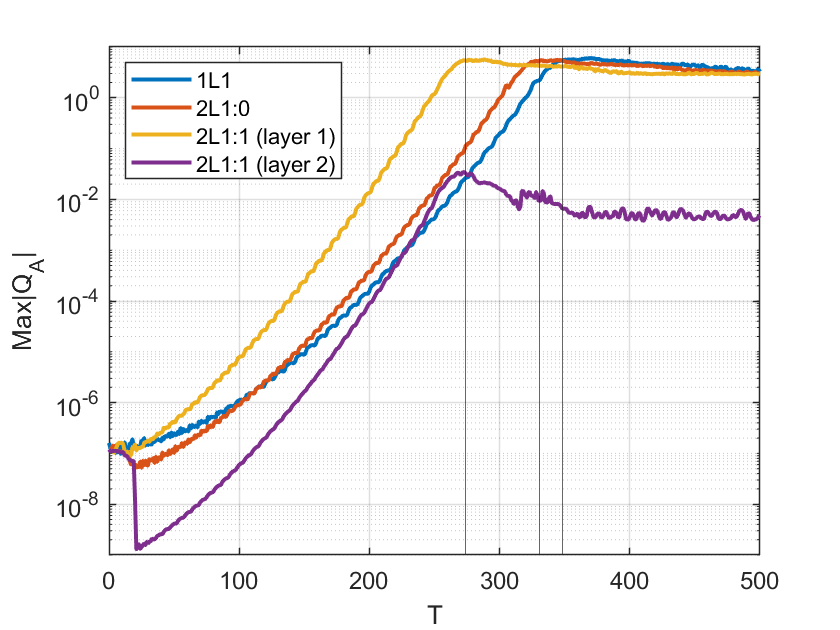}
	\end{subfigure}
	\begin{subfigure}[b]{0.24\textwidth}
        \caption{\hfill\,\vspace{-2pt}}
	\centering
	\includegraphics[trim={0cm 0.1cm 0.8cm 0.2cm},clip,width=\textwidth]{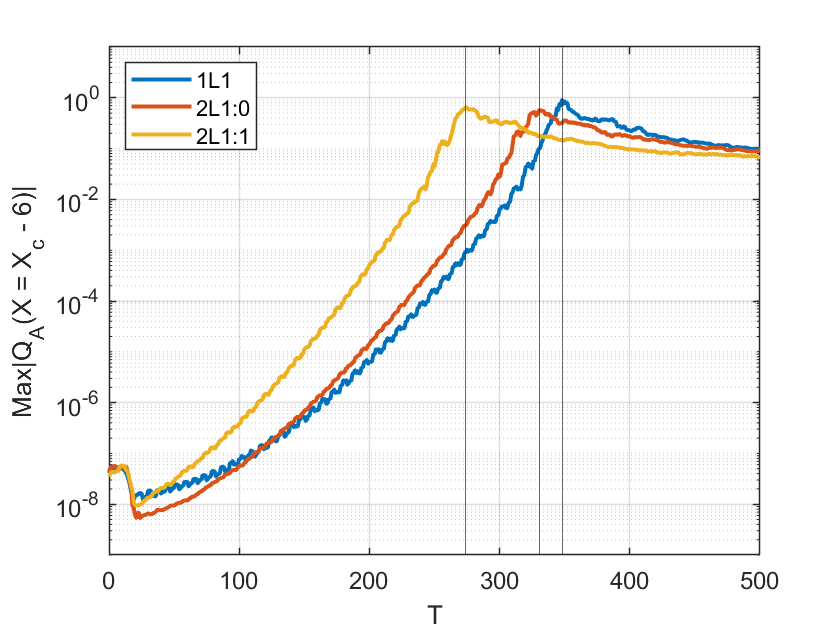}
	\end{subfigure}
 	\begin{subfigure}[b]{0.24\textwidth}
        \caption{\hfill\,\vspace{-2pt}}
	\centering
	\includegraphics[trim={0cm 0.1cm 0.8cm 0.2cm},clip,width=\textwidth]{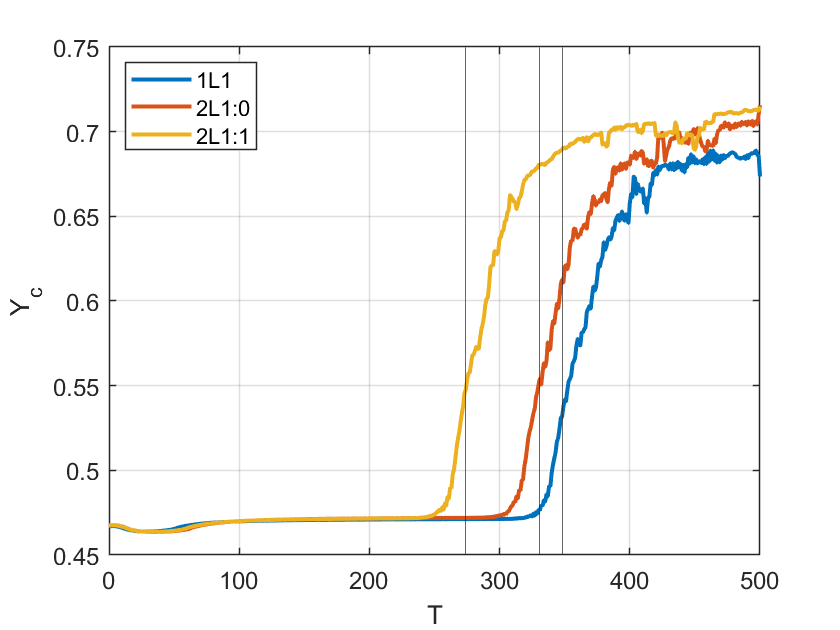}
	\end{subfigure}
	\begin{subfigure}[b]{0.24\textwidth}
        \caption{\hfill\,\vspace{-2pt}}
	\centering
	\includegraphics[trim={0cm 0.1cm 0.8cm 0.2cm},clip,width=\textwidth]{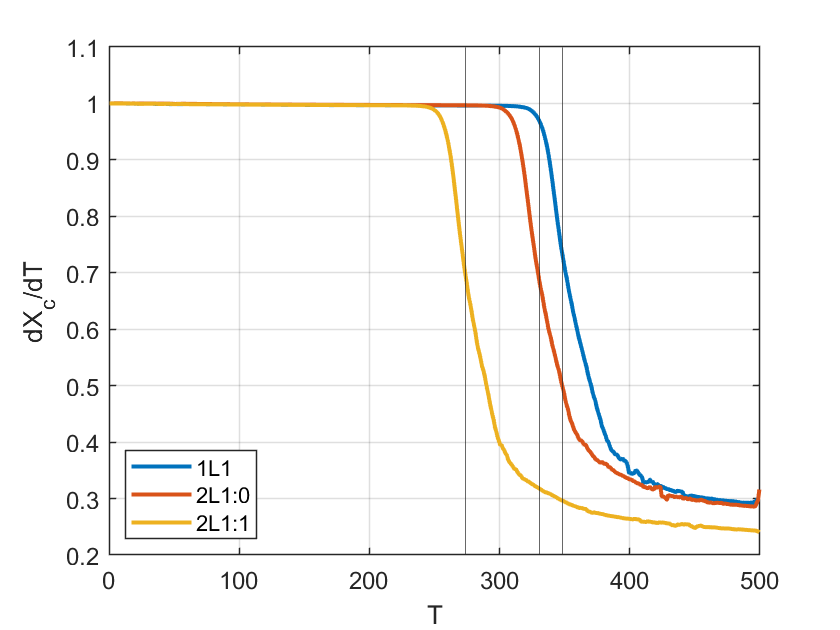}
	\end{subfigure}
	\caption{Time plots for $\beta$-plane cases: A-mode amplitude inside the dipole (a) and in the wake (b); 
    separation $Y_1$ (c) and zonal drift speed $dX_c/dT$ (d). Vertical lines denote time $T_c = 275,~330,~350$ of peak values in A-mode in the wake for three cases.}
    \label{fig:Fig_2}
 \end{figure*}


Our numerical simulations are carried out using the Julia package `GeophysicalFlows.jl' \citep{GeophysicalFlowsJOSS}, which uses GPU parallelisation to significantly reduce computational time when compared with CPU-parallelised alternatives.
Simulations use pseudo-spectral methods for spatial derivatives and a fourth-order Runge-Kutta scheme for timestepping. 
We use a doubly-periodic numerical grid of size $L_X = L_Y = 20.48$ with $2048^2$ gridpoints and run all simulations up to $T = 500$. Spectral filtering with an exponential cut-off filter \citep{SpecMethods} is used to prevent aliasing and remove the build-up of enstrophy at small scales.
Additionally, a `cutting' technique \citep{HesthavenEtAl} is applied to ensure that the periodicity does not result in the vortex interacting with it's own wake. This approach works by (smoothly) setting $Q_j = 0$ at all points further than $L_X/2-2$ from the vortex centre every $1$ time unit.

The 1- and 2-layer dipolar vortex initial conditions are determined using the Julia package `QGDipoles.jl' \citep{QGDipoles} which implements the method of \citet{Crowe_Johnson_2024} and is designed to be compatible with `GeophysicalFlows.jl'. Random initial noise of magnitude $\sim 10^{-7}$ is added to $Q_j$ in each layer.

Following \citet{Reznik} and DSB23, we analyze two parts of the computed flow field, defined by the unique decomposition $\Psi_j=\Psi_j^A+\Psi_j^S$ where
\begin{align}
\Psi_j^A &=\frac{1}{2}\left[\Psi_j(X,Y,T)+\Psi_j(X,-Y,T)\right], \\
\Psi_j^S &= \frac{1}{2}\left[\Psi_j(X,Y,T)-\Psi_j(X,-Y,T)\right].
\end{align}
Here, $\Psi_j^A$ denotes the A-component, even relative to the zonal axis, and $\Psi_j^S$ denotes the S-component, odd relative to the zonal axis.
The benefits of this decomposition are follows: the S-component represent zonal flow symmetric around the zonal axis and, 
initially, corresponds to the modon solution \cref{modon}.
Conversely, the A-component---describing the antisymmetric zonal flow perturbations---is zero initially and appears due to spontaneous symmetry breaking (DSB23).
For the multi-layer system, the equations describing nonlinear coupling of components were obtained in \citet{Sutyrin}.


We begin by considering three $\beta$-plane cases 
where $\beta_1/\beta_c = 1$. These cases are summarised in \cref{tab:tab1}. We later consider three $f$-plane cases 
where $\beta_1 = 0$ as summarized in \cref{tab:tab2}.



\cref{fig:Fig_1} shows the initial dipole spatial structure of the zonal velocities, symmetric relative the $X$-axis, and the PV, $Q_j+Y$, which has closed contours with trapped fluid only in the upper layer.
Results are shown for the 2L1:1 case, but the structure is similar in all $\beta$-plane cases with the difference in maximum zonal velocity and $Q_1$ remaining within 10\% (see \cref{tab:tab1}). 
The deep zonal velocity is smaller by approximately an order of magnitude but its pattern is wider than in the upper layer. Note that, according to the relation $Q_2 = (\beta_2/\beta_c) \Psi_2$, $Q_2$ is non-zero only in cases when $\beta_2>0$.

The symmetry breaking related to the formation of an asymmetric A-mode is characterized by time plots of its amplitude, $A_j=\max|Q_j^A|$, inside the dipole (\cref{fig:Fig_2}a) and in the Rossby wave wake defined as $\max|Q_j^A|$ on $X=X_c-6$) (\cref{fig:Fig_2}b). We plot results in the upper layer (yellow lines) and in the deep layer (magenta lines).
Here, $X_c$ denotes the $X$ position of the centre of the vortex.
We observe that the A-mode grows exponentially over time in both layers $A_j\sim \exp(\sigma T)$ for $ \sigma\approx 0.1$, saturating at around $T=T_c\simeq 275$. 
The peak values of the A-mode in the wake (\cref{fig:Fig_2}b) remains an order of magnitude smaller than inside the dipole (\cref{fig:Fig_2}a).

Nonlinear self-interaction of the A-component leads to partner separation in the S-component, $Q_j^S$, which is characterized by the meridional position of its maximum, $Y_c$, (\cref{fig:Fig_2}c).
$Y_c$ grows at a nearly constant rate, $2dY_{c}/dT \simeq 0.01$, during a phase of fast separation until $T\simeq 300$.
Corresponding weakening of the interaction between separating partners results in the reduction of the zonal drift speed $dX_c/dT$ as shown in \cref{fig:Fig_2}d.
At the time of saturation, $T=T_c$, the separation distance has increased by 15\% resulting in the zonal drift speed decreasing by 30\%. 
Later, the separation rate decreases and is accompanied by small oscillations (\cref{fig:Fig_2}c).
Correspondingly, the deceleration of the zonal drift slows for $T > 300$ (\cref{fig:Fig_2}d). 
By $T=400$, the separation increases by 60\% and the zonal drift decreases by 75\%  but remains eastward during the phase of slow separation, up until $T=500$, in all cases considered. 

\begin{figure*}
\centering
\begin{subfigure}[b]{0.42\textwidth}
\caption{\hfill\,\vspace{-2pt}}
\centering
\includegraphics[trim={0cm 0.1cm 0.1cm 0.7cm},clip,width=\textwidth]{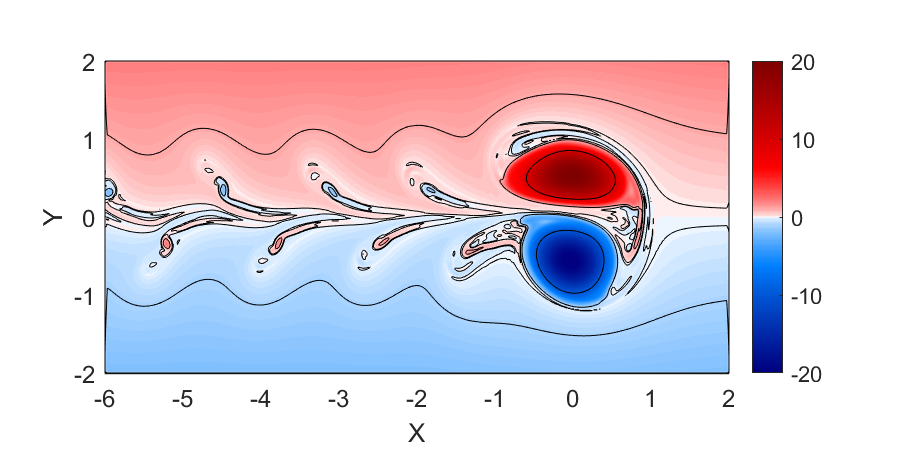}
\end{subfigure}
\begin{subfigure}[b]{0.42\textwidth}
\caption{\hfill\,\vspace{-2pt}}
\centering
\includegraphics[trim={0cm 0.1cm 0.1cm 0.7cm},clip,width=\textwidth]{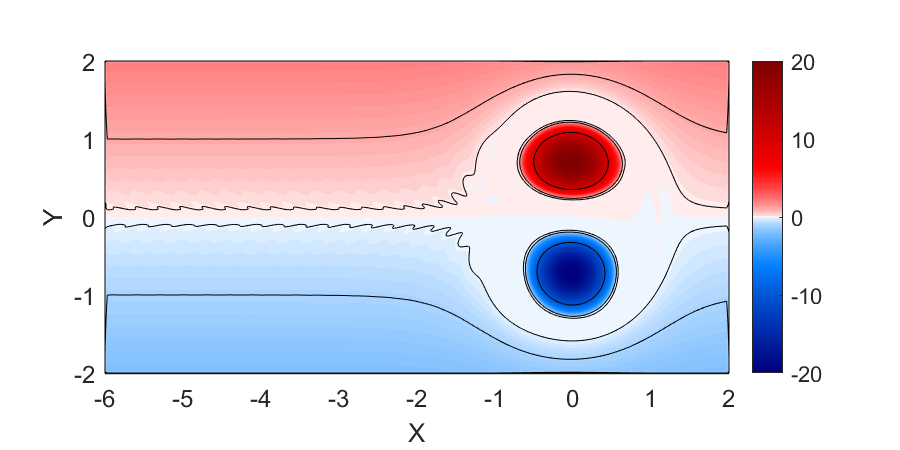}
\end{subfigure}
\begin{subfigure}[b]{0.42\textwidth}
\caption{\hfill\,\vspace{-2pt}}
\centering
\includegraphics[trim={0cm 0.1cm 0.1cm 0.7cm},clip,width=\textwidth]{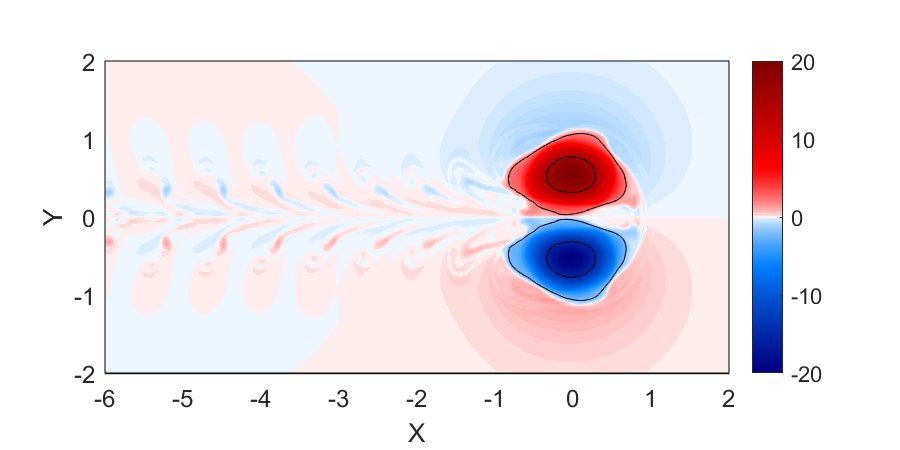}
\end{subfigure}
\begin{subfigure}[b]{0.42\textwidth}
\caption{\hfill\,\vspace{-2pt}}
\centering
\includegraphics[trim={0cm 0.1cm 0.1cm 0.7cm},clip,width=\textwidth]{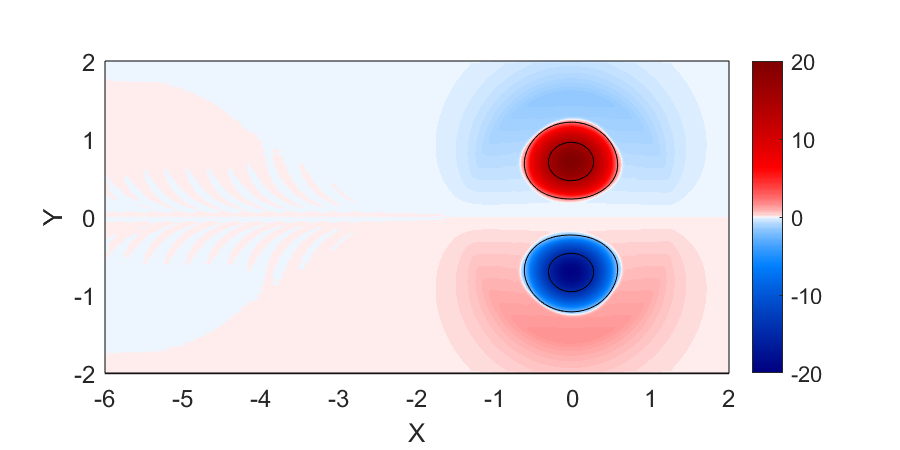}
\end{subfigure}
\begin{subfigure}[b]{0.42\textwidth}
\caption{\hfill\,\vspace{-2pt}}
\centering
\includegraphics[trim={0cm 0.1cm 0.1cm 0.7cm},clip,width=\textwidth]{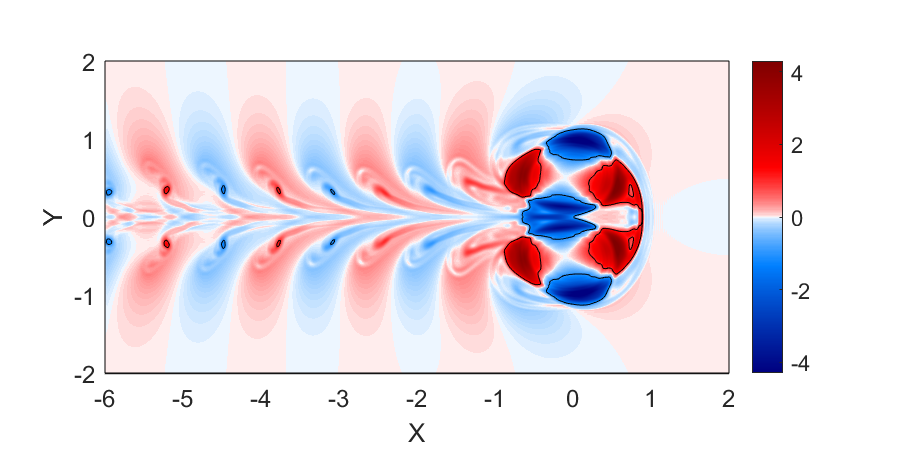}
\end{subfigure}
\begin{subfigure}[b]{0.42\textwidth}
\caption{\hfill\,\vspace{-2pt}}
\centering
\includegraphics[trim={0cm 0.1cm 0.1cm 0.7cm},clip,width=\textwidth]{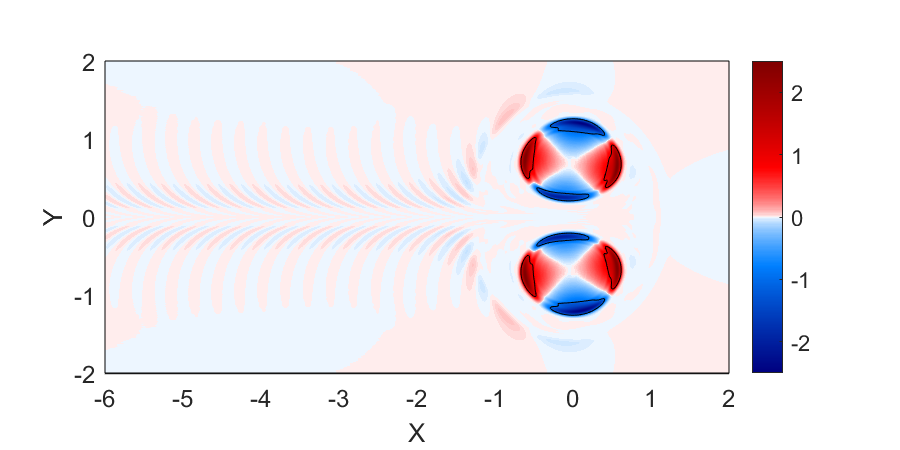}
\end{subfigure}
\begin{subfigure}[b]{0.42\textwidth}
\caption{\hfill\,\vspace{-2pt}}
\centering
\includegraphics[trim={0cm 0.1cm 0.1cm 0.7cm},clip,width=\textwidth]{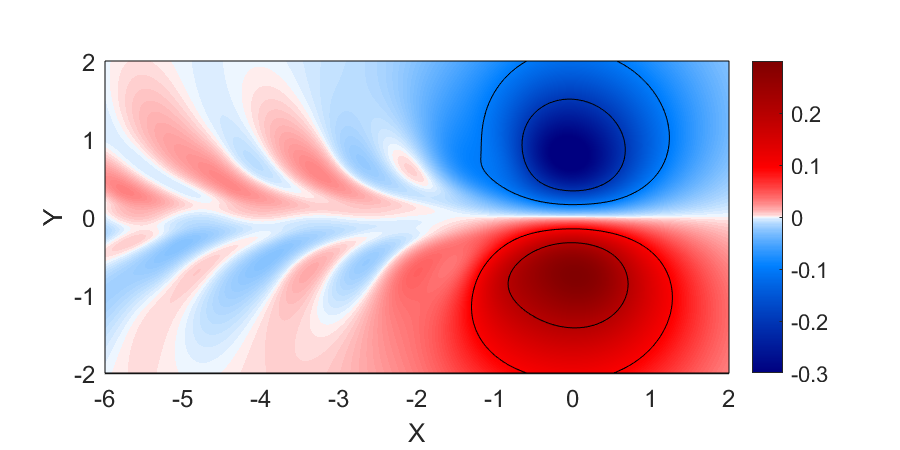}
\end{subfigure}
\begin{subfigure}[b]{0.42\textwidth}
\caption{\hfill\,\vspace{-2pt}}
\centering
\includegraphics[trim={0cm 0.1cm 0.1cm 0.7cm},clip,width=\textwidth]{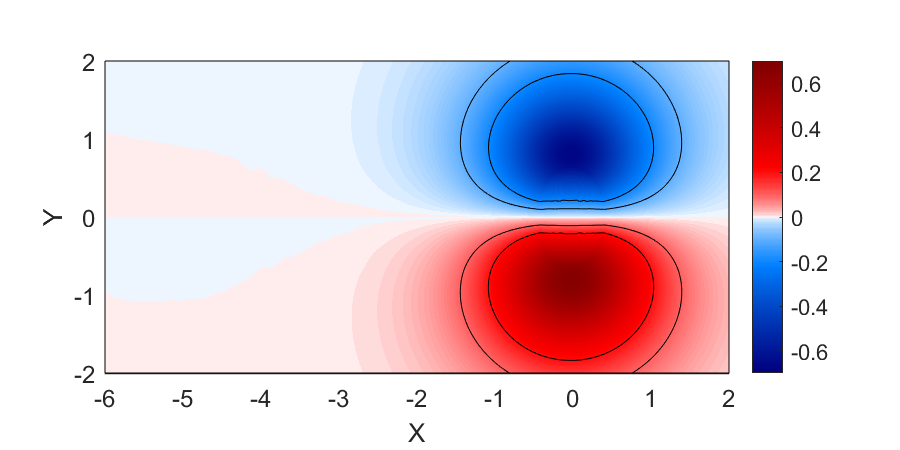}
\end{subfigure}
\begin{subfigure}[b]{0.42\textwidth}
\caption{\hfill\,\vspace{-2pt}}
\centering
\includegraphics[trim={0cm 0.1cm 0.1cm 0.7cm},clip,width=\textwidth]{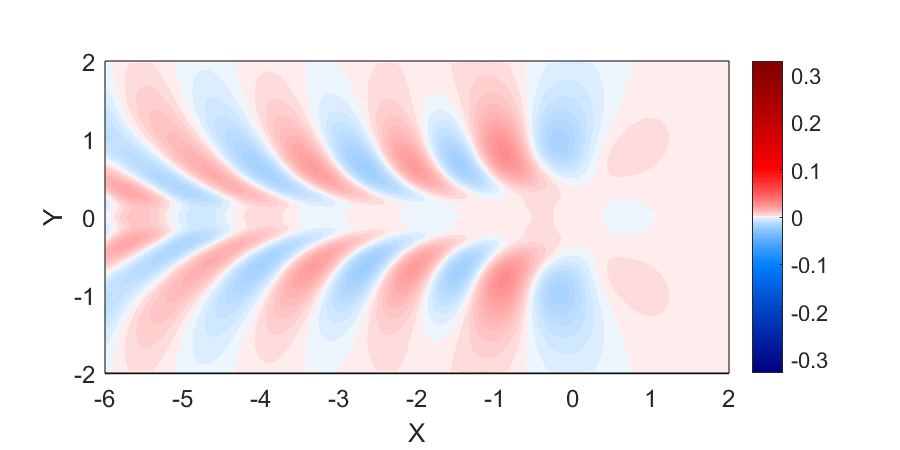}
\end{subfigure}
\begin{subfigure}[b]{0.42\textwidth}
\caption{\hfill\,\vspace{-2pt}}
\centering
\includegraphics[trim={0cm 0.1cm 0.1cm 0.7cm},clip,width=\textwidth]{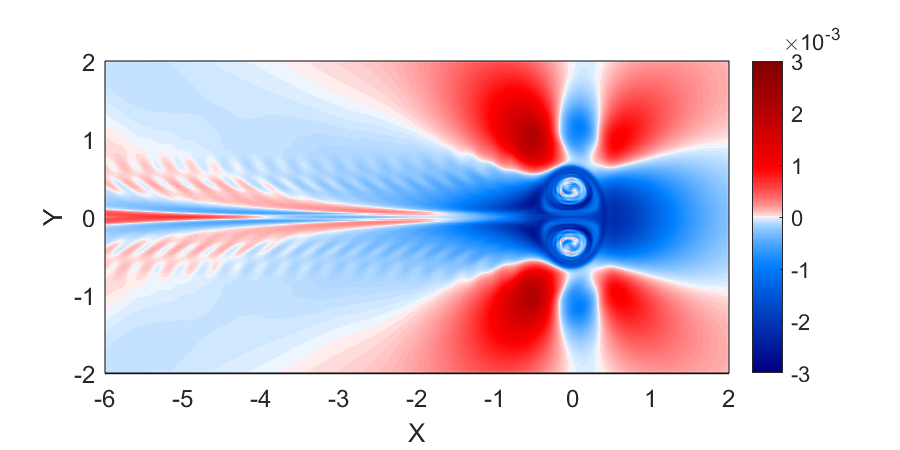}
\end{subfigure}
\caption{The snapshots of $Q_1+Y,~Q_1^S,~ Q_1^A,~ Q_2,~ Q_2^A$ for case 2L1:1 at $T\simeq T_c$ (left) and $T\simeq 500$ (right). The phases are chosen for better comparison to Figure 2d in \citet{DaviesEtAl_2}. Supplementary movies files, Mov\_1.mp4 - Mov\_5.mp4, show the evolution of these fields over $T\in[200, 500]$.}
\label{fig:Fig_3}
 \end{figure*}

\begin{figure*}
	\centering
	\begin{subfigure}[b]{0.24\textwidth}
        \caption{\hfill\,\vspace{-2pt}}
	\centering
	\includegraphics[trim={0.0cm 0.1cm 0.8cm 0.2cm},clip,width=\textwidth]{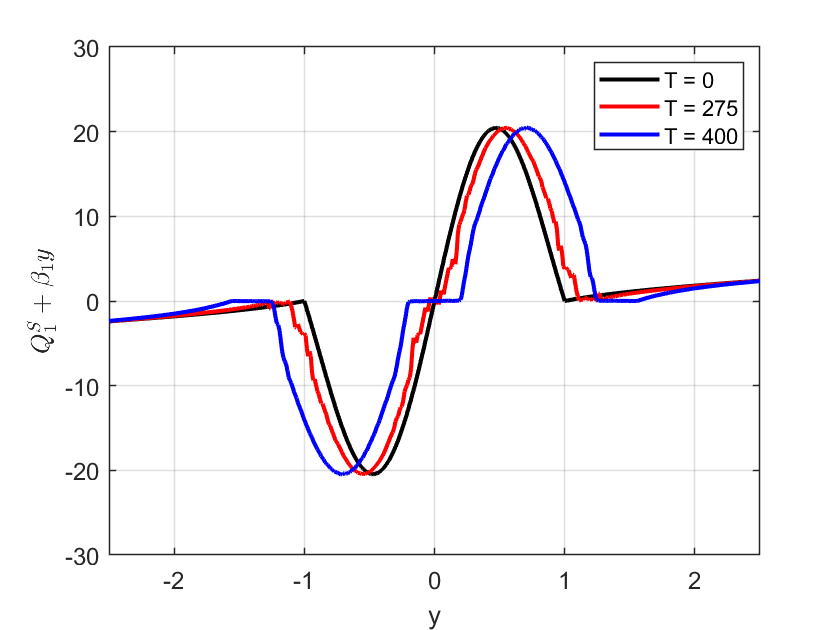}
	\end{subfigure}
	\begin{subfigure}[b]{0.24\textwidth}
        \caption{\hfill\,\vspace{-2pt}}
	\centering
	\includegraphics[trim={0.0cm 0.1cm 0.8cm 0.2cm},clip,width=\textwidth]{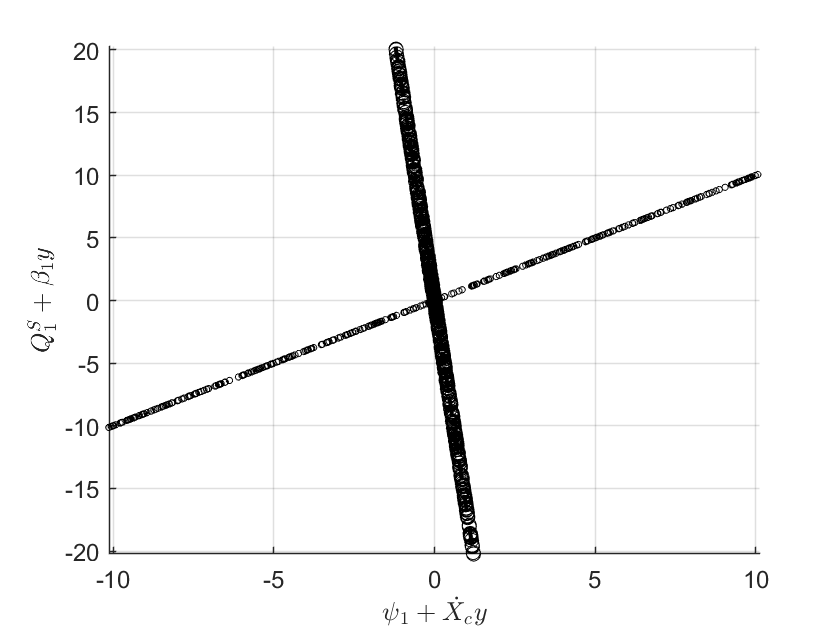}
	\end{subfigure}
	\begin{subfigure}[b]{0.24\textwidth}
        \caption{\hfill\,\vspace{-2pt}}
	\centering
	\includegraphics[trim={0.0cm 0.1cm 0.8cm 0.2cm},clip,width=\textwidth]{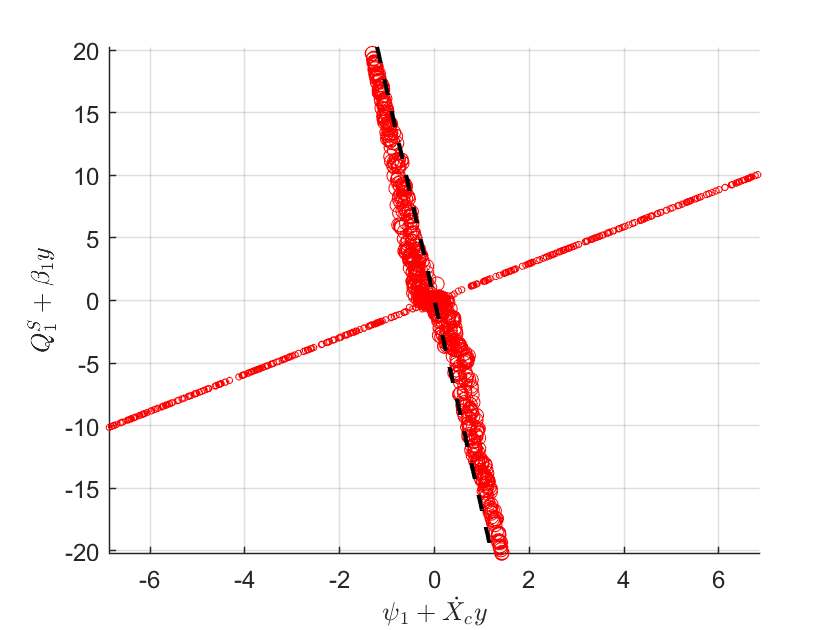}
	\end{subfigure}
	\begin{subfigure}[b]{0.24\textwidth}
        \caption{\hfill\,\vspace{-2pt}}
	\centering
	\includegraphics[trim={0.0cm 0.1cm 0.8cm 0.2cm},clip,width=\textwidth]{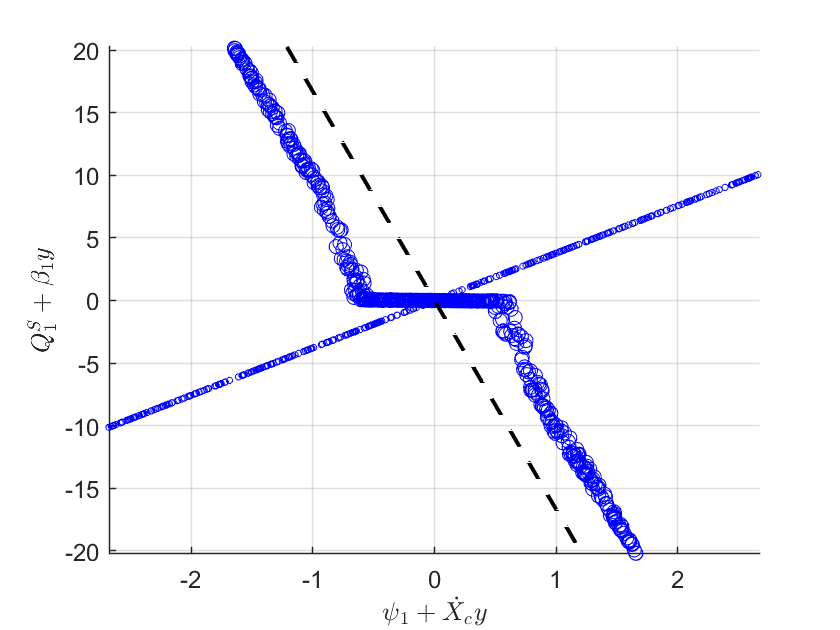}
	\end{subfigure}
	\caption{Meridional sections of $Q_1^S(X_c,Y,T)+\beta_1Y$ (a) and scatter plots of $Q_1^S+\beta_1Y$ vs 
    $\Psi_1^S+\dot X_cY$ at the same times $T=0$ (black) (b); $T=T_c$ (red) (c); $T=400$ (blue) (d). All results shown for the 2L1:1 case.}
    \label{fig:Fig_4}
 \end{figure*}

Figure 3 shows snapshots from supplemental animations for case 2L1:1 (Mov\_1.mp4 - Mov\_5.mp4)
near the saturation, $T\simeq T_c$ (left panels), and at $T\simeq 500$ (right panels).
Growing rotating asymmetries in the upper layer (\cref{fig:Fig_3}e) are superimposed on the separating partners in the S-component (\cref{fig:Fig_3}c) resulting in elongation and compression of the vortex pair (\cref{fig:Fig_3}a) similar to observations in DSB23. 
Rossby waves are also seen in the lower layer (\cref{fig:Fig_3}g, i).
Small scale details in the wake become visible when approaching saturation (\cref{fig:Fig_3}e) and appear to be related to  oscillations of a hyperbolic point behind the dipole core (\cref{fig:Fig_3}a), resulting in the shedding of core material as seen in movies.
Eventual generation of Rossby waves in the S-component (\cref{fig:Fig_3}c) results from nonlinear self-interaction of the A-mode.
After saturation, the pattern of the A-mode gradually changes during partner separation and deceleration and we observe a substantial weakening of the Rossby wave radiation, accompanied by a shortening of the wavelength (\cref{fig:Fig_3}b, d, f, h, j).
Note that a meridionally elongated vortex core forms through the homogenization of PV between the oscillating partners and the separatrix (\cref{fig:Fig_3}b).
This can be further seen by comparing instantaneous meridional sections of $Q_1^S(X_c,Y)+\beta_1Y$ with the initial section as shown in \cref{fig:Fig_4}a. 
In accordance with the Lagrangian conservation of PV extrema, we observe that the maximum PV remains nearly constant as the vortex centers move outwards, resulting in
the S-component weakening by only 2\%. 
Apparently, this small weakening of the partners allows for a long-lasting phase of slow partner separation and oscillations with the radiation of short Rossby waves (\cref{fig:Fig_3}d, f, j). 

In order to show that $Q_1^S$ tends to a quasi-steady state (\cref{fig:Fig_3}d), scatter plots of $Q_1^S+\beta_1Y$ vs $\Psi_1^S+\dot X_c Y$ are shown for $T=0$ (\cref{fig:Fig_4}b), for $T=T_c$ (\cref{fig:Fig_4}c) and for $T=400$ (\cref{fig:Fig_4}d).
Each circle corresponds to a point on the numerical grid with larger circles denoting points in $r <1$ and smaller circles denoting points in $r>1$.
A black dashed line shows $\mathcal{F}_1(z) = -\kappa_1^2\,z$, the relationship initially satisfied for $r<1$.
At later times, we observe a tendency to form nonlinear $\mathcal{F}_1(z)$, typical of the non-circular dipoles.
The points in $r > 1$ are consistent with the linear relationship $\mathcal{F}_1(z) = (\beta a^2 / \dot{X}_c)\,z$ throughout the evolution.
Thus, the symmetry breaking results in a nonlinear transformation from the initial steady state to a pulsating EPD with an oscillating A-mode and slowly evolving S-component, $Q_1^S$.
This regime is not observed for larger $\beta$ values where the dipole disintegrates after the EPD separate into two monopolar vortices drifting in a westward direction (DSB23). 

\begin{table*}
\centering
\begin{tabular}{ccccc} 
 Case & two-layer sloping bottom & two-layer flat bottom  & upper layer \\
 Notation & 2L0:1 & 2L0:0 & 1L0 \\
 $\beta_2/\beta_c$ & 1 & 0 & 0 \\
 $U_{1max}$ & 4 & 4 & 4.5 \\
 $Q_{1max}$ & 16 & 15 & 17.5 \\
\end{tabular}
\caption{Summary of the $f$-plane cases.}
\label{tab:tab2}
\end{table*}

\begin{figure*}
	\centering
	\begin{subfigure}[b]{0.24\textwidth}
        \caption{\hfill\,\vspace{-2pt}}
	\centering
	\includegraphics[trim={0.0cm 0.1cm 0.8cm 0.2cm},clip,width=\textwidth]{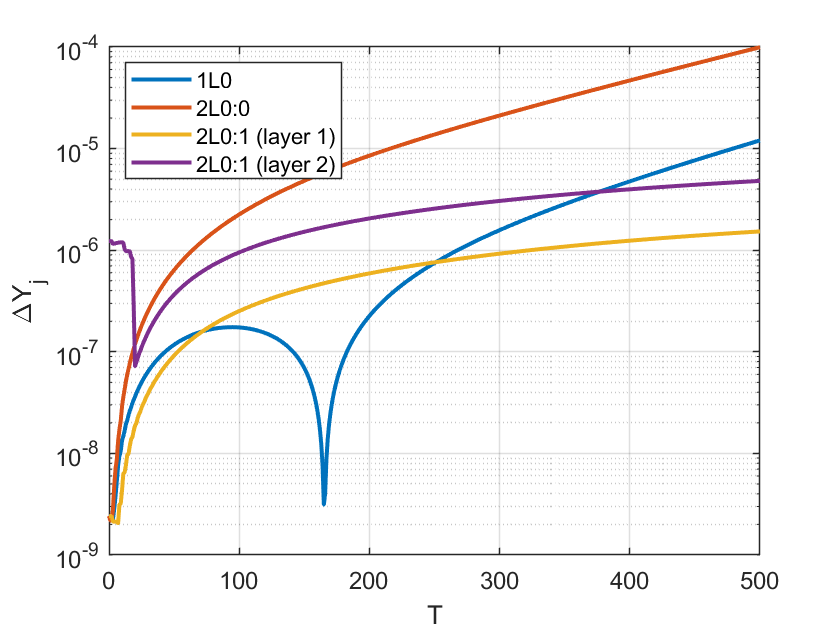}
	\end{subfigure}
	\begin{subfigure}[b]{0.24\textwidth}
        \caption{\hfill\,\vspace{-2pt}}
	\centering
	\includegraphics[trim={0.0cm 0.1cm 0.8cm 0.2cm},clip,width=\textwidth]{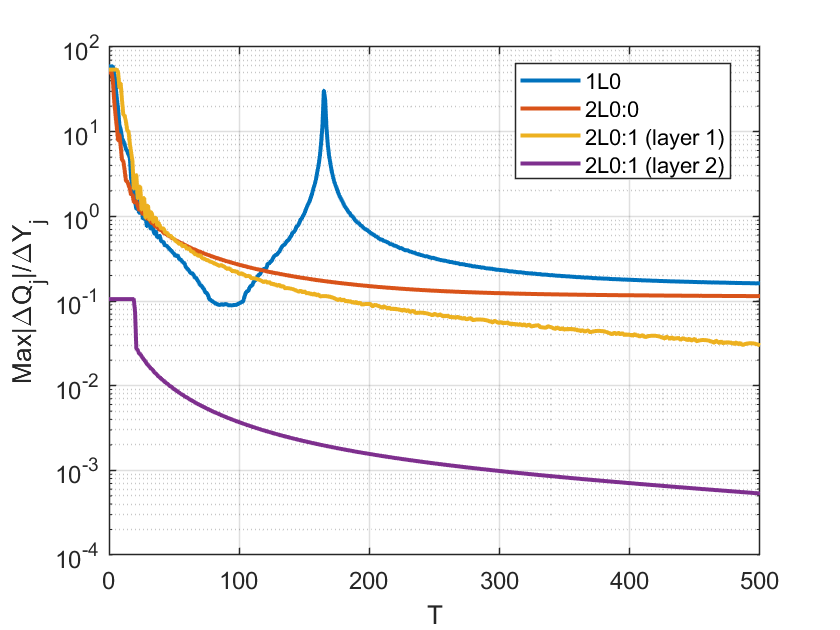}
	\end{subfigure}
	\begin{subfigure}[b]{0.24\textwidth}
        \caption{\hfill\,\vspace{-2pt}}
	\centering
	\includegraphics[trim={0.0cm 0.1cm 0.8cm 0.2cm},clip,width=\textwidth]{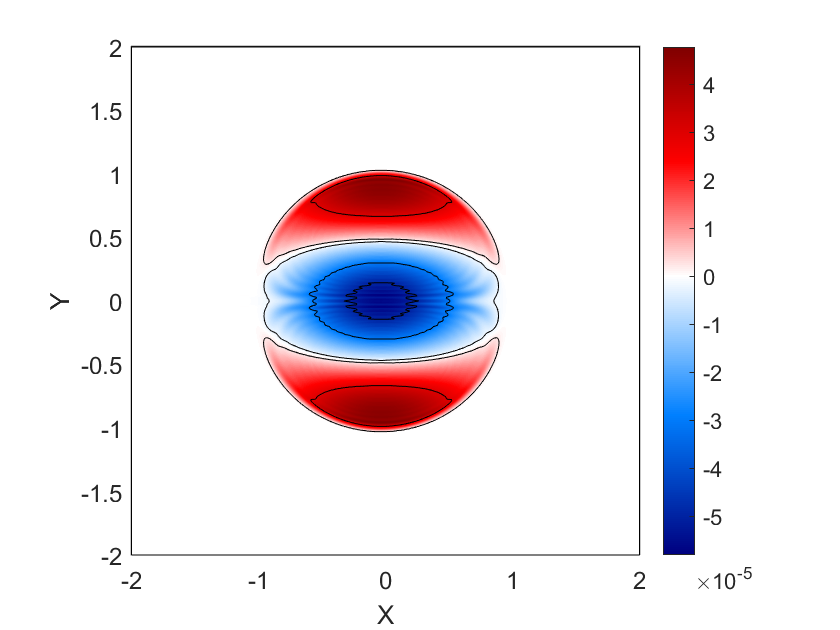}
	\end{subfigure}
	\begin{subfigure}[b]{0.24\textwidth}
        \caption{\hfill\,\vspace{-2pt}}
	\centering
	\includegraphics[trim={0.0cm 0.1cm 0.8cm 0.2cm},clip,width=\textwidth]{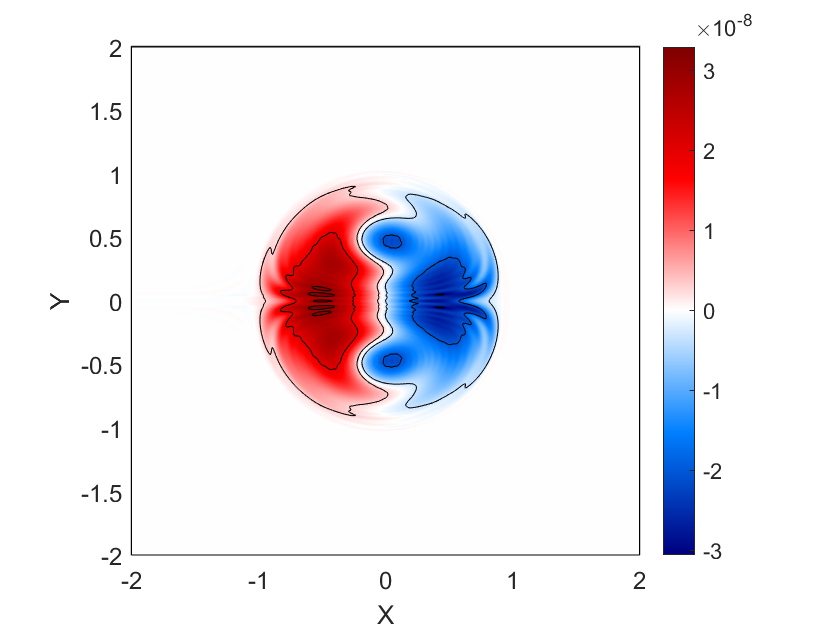}
	\end{subfigure}
	\caption{Plots for the three $f$-plane cases: time plots of the meridional shift evaluated by \cref{eq:DeltaY} (a) and the relative amplitude of residual field given by \cref{eq:DeltaQ} (b); snapshots of the A-mode (c) and the residual field (d) in the upper layer at $T\simeq 500$ for the 2L0:1 case.}
    \label{fig:Fig_5}
 \end{figure*}

Similar behaviour is found in the case 2L1:0 which corresponds to the `quasi-reduced-gravity model' where $Q_2$ remains zero due to an exact compensation of $\beta$ by the topographic slope in the lower layer \citep{KiznerEtAl}.
Again we see symmetry breaking due to the development of an A-mode while the lack of a PV gradient in the lower layer retards the A-mode growth (\cref{tab:tab1}) and saturation (\cref{fig:Fig_2}a, b), resulting in slightly reduced separation (\cref{fig:Fig_2}c) and deceleration (\cref{fig:Fig_2}d).


Next we consider case 1L1 where we simulate only the dynamics of the upper layer in the reduced-gravity model, neglecting lower layer feedback.
For an initial Larichev-Reznik EPD, the A-mode grows slightly slower than for the two-layer cases (\cref{tab:tab1}) and its spatio-temporal structure looks similar to that found in DSB23 for $\beta > \beta_c$ where the time of integration was limited ($T < 60$).
The more efficient code here allows for longer integration, allowing us to evaluate the growth rate $\sigma \simeq 0.06$ (\cref{fig:Fig_2}a) and observe saturation at $T_c=350$.
In addition to the weakly nonlinear effects analyzed in \citet{DaviesEtAl_3}, we see a transition in A-mode structure from the fast separation phase into a new pattern during slow separation, similar to the case 2L1:1 analyzed above (see \cref{fig:Fig_3} and supplemental animations).
Running with $\beta_1=\beta_c/2$ results in half the grown rate, $\sigma \simeq 0.03$, indicating its proportionality to the value of $\beta$ and implying a lack of critical $\beta/\beta_c$ for spontaneous symmetry breaking.


Drastically different behaviour is found in three $f$-plane cases 
where $\beta_1 = 0$. These cases are summarised in \cref{tab:tab2} and correspond to cases where a PV gradient can exist owing to sloping topography.
Here we do not see symmetry breaking as the A-mode pattern (\cref{fig:Fig_5}c) is well approximated by a meridional shift of the vortex center, proportional to the initial meridional gradient of $Q_j^S$. 
\cref{fig:Fig_5}a shows the evolution of the shift defined by  
\begin{equation}
\label{eq:DeltaY}
\Delta Y_j(T)=\max|Q_j^A|[dQ_j^S/dY(X_c,0)]^{-1},
\end{equation}  
showing its decreasing growth rate with time and overall small magnitude, $\Delta Y_j(T) \simeq 10^{-5}$. 
\cref{fig:Fig_5}b shows the relative amplitude of the residual field, $\max|\Delta Q_1|/\Delta Y_1$, where
\begin{multline}
\label{eq:DeltaQ}
\Delta Q_j=Q_{j}^A+Q_j^S(X-X_c, Y-\Delta Y_j)+Q_j^S(X-X_c, Y) \\ \approx
Q_j^A-\Delta Y_j\frac{dQ_j^S}{dY}(X-X_c,Y). 
\end{multline}
In the case 2L0:1, the residual field (\cref{fig:Fig_5}d) tends to three orders of magnitude smaller than the A-mode amplitude (\cref{fig:Fig_5}c), suggesting that the asymmetry can be well explained by a meridional shift in the vortex, rather than the formation of an A-component.
Therefore, despite the presence of the topographic $\beta$-effect in the lower layer, the evolution remains nearly symmetric in contrast to an exponentially growing A-mode seen for cases with a PV gradient in the upper layer.

For the case 2L0:0, the dipole remains steady propagating with a slightly larger meridional shift $\simeq 10^{-4}$ (red line in \cref{fig:Fig_5}b), than 2L1:0 where $\beta_2>0$ in the lower layer (yellow line in \cref{fig:Fig_5}b). 

Finally, in contrast to the two-layer dipoles, in the upper layer only case, 1L0, we see a sharp decrease in the meridional shift at $T\simeq 170$ (blue line in \cref{fig:Fig_5}b) related to a change in sign. The subsequent growth and decrease in the residual field is similar to the other $f$-plane cases.


In summary, we have studied the evolution of eastward propagating dipoles (EPD) in a two-layer, quasi-geostrophic, $\beta$-plane model using high-resolution numerical simulations. Including a sloping bottom allows for various combinations of the background PV gradient, $\beta_j$, in the upper ($j=1$) and lower ($j=2$) layers.

In the cases of a non-zero $\beta_1$, corresponding to a PV gradient in the upper layer with an initial dipolar PV anomaly, the symmetry of the dipole flow breaks due to an exponentially growing, rotating, asymmetric A-mode of linear instability associated with Rossby wave radiation.
The nonlinear self-interaction of this growing A-mode results in fast partner separation and deceleration of the easward drift. 
Further, a regime of slowly separating, oscillating partners with a saturated A-mode and weaker and shorter Rossby waves is revealed for $\beta_1 \le \beta_c$. In this regime the S-component approaches a quasi-steady state with a nonlinear $\mathcal{F}_1(z)$ corresponding to a meridionally elongated core. 
Such long-lasting pulsating EPD display new behaviours which have not previously been observed.

In the case of no $\beta$-effect in the upper layer, the dipoles remains nearly symmetric even when a PV gradient is present in the lower layer. 
The asymmetric mode which appears in these configurations is well described by small meridional shift of the dipole center.

Our results shed new light on the rich dynamics of dipolar vortices in the two-layer setup and recently discovered breakdown mechanisms in one layer. Additionally, this work highlights the effectiveness of GPU-parallelized code in the study of long-lived phenomena.
Given the limited scope of this short paper, we present only a few examples of dipole instability and nonlinear transformation. 
As such, further work is required to fully understand the physical mechanisms behind these processes. 
In particular, our numerical results indicate a possibility of exploring the instability semi-analytically.
Additionally, the assumptions placed on the layer depths ($h_1 = h_1$) and vortex radius ($a = R_1$) restrict our conclusions, hence a broader study of parameter space is required to understand the parameter dependence of these dynamical processes.

\vspace{0.3cm}
\noindent
\textbf{Supplementary Material}

See the supplementary material for animations of the dipole evolution that support the results of this study.

\vspace{0.3cm}
\noindent
\textbf{Acknowledgments}

The authors thank Prof. Jonas Nycander for useful comments on animating the dipole evolution, and Prof. Ted Johnson for helpful discussions on dipole instabilities.

\vspace{0.3cm}
\noindent
\textbf{AUTHOR DECLARATIONS}

\noindent
\textbf{Conflict of Interest}

The authors have no conflicts to disclose.

\vspace{0.3cm}
\noindent
\textbf{Author Contributions}

\noindent
\textbf{Matthew N. Crowe:} Conceptualization (equal); Data curation (lead); Formal
analysis (equal); Investigation (equal); Methodology (equal); Visualization (lead); Writing – review \& editing (equal).

\noindent
\textbf{Georgi G. Sutyrin:}
Conceptualization (equal); Methodology (equal); Supervision (equal);
Writing – original draft (lead).

\vspace{0.3cm}
\noindent
\textbf{DATA AVAILABILITY}

The data that support the findings of this study are available from the corresponding author upon reasonable request.

\vspace{0.3cm}
\noindent
\textbf{REFERENCES}
\bibliography{bibliography}

\end{document}